\documentclass[journal]{IEEEtran}

\ifCLASSINFOpdf
\else
   \usepackage[dvips]{graphicx}
\fi
\usepackage{url}

\usepackage[UTF8, scheme = plain, fontset=windows]{ctex}
\usepackage{booktabs}
\usepackage{multirow}
\usepackage{graphicx}
\usepackage{wrapfig}
\usepackage{subfigure}
\usepackage{makecell}
\usepackage{indentfirst}
\usepackage[linesnumbered,ruled]{algorithm2e}
\usepackage{algpseudocode}
\usepackage{epstopdf}
\usepackage[table,xcdraw]{xcolor}
\usepackage{color}
\usepackage{amssymb}
\usepackage{amsmath}
\usepackage{titlesec}
\usepackage{cite}
\usepackage{arydshln}

\begin{document}

\title{Protecting Your Voice: \\ Temporal-aware Robust Watermarking}

\author{Yue Li, Weizhi Liu, Dongdong Lin, Hui Tian, \IEEEmembership{Senior Member, IEEE}, Hongxia Wang, \IEEEmembership{Member, IEEE}
%\thanks{Yue Li, Weizhi Liu, and Hui Tian are with the College of Computer Science and Technology, Huaqiao University, Xiamen 361021, China, and also with the Xiamen Key Laboratory of Data Security and Blockchain Technology, Xiamen 361021, China.}\textit{Corresponding author: Yue Li and Hui Tian.}
\thanks{Yue Li, Weizhi Liu, Dongdong Lin and Hui Tian are with the College of Computer Science and Technology, National Huaqiao University, Xiamen 361021, China, and also with the Xiamen Key Laboratory of Data Security and Blockchain Technology, Xiamen 361021, China (e-mail: liyue\_0119@hqu.edu.cn; lwzzz@stu.hqu.edu.cn; dongdonglin8@gmail.com;
htian@hqu.edu.cn).

Hongxia Wang is with the School of Cyber Science and Engineering, Sichuan University, Chengdu 610207, China (e-mail:hxwang@scu.edu.cn)
}
\thanks{\textit{Corresponding author: Hui Tian}}
}

\markboth{Journal of \LaTeX\ Class Files, December 2024}
{Shell \MakeLowercase{\textit{et al.}}: Bare Demo of IEEEtran.cls for IEEE Journals}
\maketitle

\begin{abstract}
The rapid advancement of generative models has led to the synthesis of real-fake ambiguous voices. To erase the ambiguity, embedding watermarks into the frequency-domain features of synthesized voices has become a common routine. However, the robustness achieved by choosing the frequency domain often comes at the expense of fine-grained voice features, leading to a loss of fidelity.
Maximizing the comprehensive learning of time-domain features to enhance fidelity while maintaining robustness, we pioneer a \textbf{\underline{t}}emporal-aware \textbf{\underline{r}}ob\textbf{\underline{u}}st wat\textbf{\underline{e}}rmarking (\emph{True}) method for protecting the speech and singing voice.  
%through comprehensive learning of time-domain features. 
%To address this limitation, we meticulously designed a structure-lightweight 
%encoder tailored for reconstructing the waveform with the watermark embedded in time-domain features. Moreover, we redesigned a temporal-aware gated convolution network for bit-wise watermark recovery.
%Comprehensive experiments and comparisons with existing state-of-the-art methods have demonstrated the vigorous robustness and superior fidelity of the proposed approach achieving an average PESQ score of 4.63.
For this purpose, the integrated content-driven encoder is designed for watermarked waveform reconstruction, which is structurally lightweight. Additionally, the temporal-aware gated convolutional network is meticulously designed to bit-wise recover the watermark. 
Comprehensive experiments and comparisons with existing state-of-the-art methods have demonstrated the superior fidelity and vigorous robustness of the proposed \textit{True} achieving an average PESQ score of 4.63.
\end{abstract}


\IEEEpeerreviewmaketitle

% -----------------------------------------------------------
\vspace{-0.3cm}
\section{Introduction}
\label{sec_intro}

\IEEEPARstart{G}{enerative} models have significantly advanced text-to-speech and text-to-music synthesis technologies~\cite{tan2024naturalspeech,kim2021vits,zhang2024stylesinger,liu2022diffsinger}, breaking the high-tech barrier of voice cloning technology. With their ability to closely mimic voices, these innovations raise increasing concerns, particularly in facilitating events such as fraud~\cite{deepfake_audio_fraud} and misinformation campaigns \cite{deepfake_voice}. In response, governments and regulatory bodies are developing policies (CHN~\cite{CHN_POLICY}, EU~\cite{EU_POLICY} and USA~\cite{USA_POLICY}) aimed at regulating AI-generated content (AIGC) to mitigate these risks. 
%enabling the generated waveform that is arduous to differentiate from natural singing voice and speech voice. 

%Without proper control mechanisms, such capabilities could culminate in the proliferation of deepfake waveform, potentially infringing upon the intellectual property rights of natural voice. In this scenario, a class of methods has emerged that utilizes the transferability of the watermark from training data (natural voice) to generative models for proactive forensic~\cite{yu2021artificial, fernandez2023stable, jang2024waterf}. And the authorized natural voice is a critical component that significantly contributes to the efficacy of this method~\cite{wang2024diagnosis}. To this front, leveraging deep-learning-based watermarking techniques to protect the natural voice becomes a crucial aspect of these transferability-based methods.

All of the aforementioned regulatory policies emphasize the responsibility of AIGC companies to implement specific marks for generated content. This underscores the significance of \textit{watermarking} technology as an essential solution for proactive regulation of deepfake content \cite{yu2021artificial, yuresponsible, fernandez2023stable, asnani2024promark}. As a consequence, research has increasingly shifted from traditional handcrafted watermarking methods to deep-learning-based approaches in the field of multimedia watermarking. 

\begin{figure}
    \centering
    \setlength{\abovecaptionskip}{-0.01mm}
    \setlength{\belowcaptionskip}{-0.1mm}
    \resizebox{0.95\linewidth}{!}{\includegraphics{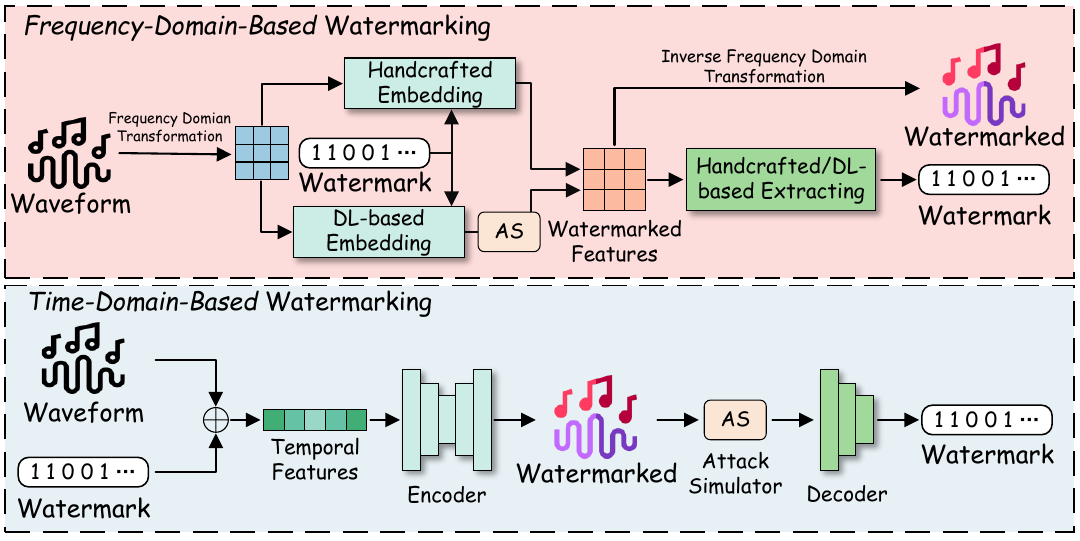}}
    \caption{Two branches of audio watermarking methods. Upper: Frequency-domain watermarking dominates existing SOTA methods. Lower: The proposed temporal-aware watermarking preserves robustness without compromising fine-grained temporal features.}
    \label{fig_overview}
\vspace{-0.7cm}
\end{figure}

In the field of audio and voice watermarking, it has seemingly become an established principle that achieving good robustness necessitates a frequency domain transformation (FDT). As depicted in the upper branch of Fig. \ref{fig_overview}, this holds true both in the earlier era of handcrafted watermarking~\cite{zhang2023m_sw_lsc, zhao2022ssvs, zhao2021desyn-fsvcm, saadi2019normspace, liu2018patchwork, natgunanathan2017patchwork} and in the current epoch dominated by deep-learning-based approaches~\cite{deepmind2024synthid, roman2024audioseal, liu2023timebre, liu2023dear, qu2023audioqr, chen2023wavmark, pavlovic2022robust,or2024maskmark}.
Concretely, Liu et al.~\cite{liu2023dear} embedded the watermark into the approximate coefficients derived from the Discrete Wavelet Transform (DWT). Fully leveraging the short time window properties of the Short-Time Fourier Transform (STFT), Liu et al.~\cite{liu2023timebre} employed frequency features for robust watermarking. 
Additionally, the advantageous properties of STFT magnitudes have motivated subsequent works, including Chen et al.~\cite{chen2023wavmark}, Pavlovic et al.~\cite{pavlovic2022robust}, and O'Reilly et al.~\cite{or2024maskmark}, to embed watermarks into these magnitudes. For DL-based approaches mentioned above, another notable observation is that achieving high robustness requires more than just FDTs---the use of an attack simulator (AS) is essential. 
This observation prompts us to rethink: \textit{Between FDT and AS, which serves as the cornerstone of robustness? Can robust performance be achieved with the AS alone, in the absence of FDT?} If FDT proves non-essential, it may be feasible to shift the watermark embedding process from the frequency domain to the temporal domain, potentially enabling fine-grained manipulation of timbre features for improved fidelity.

\begin{figure*}
    \centering
    \setlength{\abovecaptionskip}{-0.01cm}
    \setlength{\belowcaptionskip}{-0.01cm}
    \resizebox{0.87\textwidth}{!}{\includegraphics{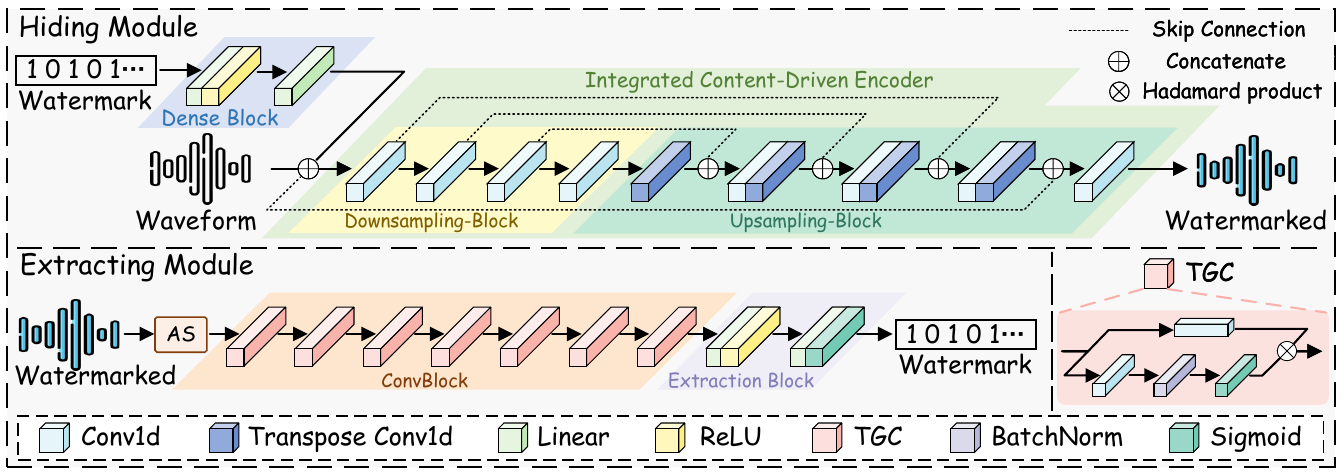}}
    \caption{\textbf{Detailed Architecture of True.} The hiding module consists of a dense block and an integrated content-driven encoder for watermarking. The extracting module includes the attack simulator (AS), the convolutional block (ConvBlock), and an extraction block. The ConvBlock is composed of several temporal-aware gated convolution (TGC) networks.}
    \label{fig_archi}
\vspace{-0.6cm}
\end{figure*}

To validate the feasibility of the aforementioned assumption, we propose a  \underline{t}emporal-aware and \underline{r}ob\underline{u}st wat\underline{e}rmarking (\textbf{True}) method tailored for speech and singing voices. The proposed method directly embeds the watermark into the voice waveform, as illustrated in the lower branch of Fig.~\ref{fig_overview}, thereby avoiding the timbre detail loss typically associated with frequency-domain transformations in conventional frequency-domain-based watermarking approaches. Meanwhile, the AS has borne the responsibility of ensuring the watermark's robustness.  Thus, the contributions can be boiled down to:

\begin{itemize}
    \item \textit{New Paradigm.} We pioneer a novel temporal-aware watermarking method, \textit{True}, designed to proactively protect the copyright of singing and speech voices, with only AS needed to guarantee the robustness of this paradigm.

    \item \textit{Novel Architecture.} To preserve fine-grained voice features, an integrated content-driven encoder is proposed for watermarking. For bit-wise extraction, a temporal-aware gated convolutional network (TGC) is designed. 

    \item \textit{Sound Performance.} Comprehensive experiments and comparisons with state-of-the-art (SOTA) methods demonstrate the superior fidelity and enhanced robustness.
\end{itemize}

% -----------------------------------------------------------
\vspace{-0.6cm}
\section{Methodology}
\label{sec_method}
The proposed method, \textit{True}, aims to directly leverage the time-domain features from the voice waveform, achieving the trade-off between robustness and fidelity for the watermarked waveform. To this end, the method encompasses two main components: the hiding module (HM) and the extracting module (EM), as illustrated in Fig.~\ref{fig_archi}. All components are jointly trained through an optimization strategy, the details of which are elaborated below.
%The proposed voice watermarking method, as illustrated in Fig.~\ref{fig_archi}, includes a hiding module (HM), an extraction module (EM), and an attack simulator (AS). Concretely, the HM leverages the time-domain features of the waveform for watermarking, enabling direct reconstruction of the watermarked waveform. The EM extracts the watermark more precisely from the fine-grained time-domain features of the waveform. Afterward, the AS incorporates common voice post-processing operations to bolster the robustness. All components undergo training through a joint optimization strategy, the details of which will be elaborated as follows.
% -----------------------------------------------------------
\vspace{-0.3cm}
\subsection{Hiding Module}
\label{sec_hm}
In the hiding module, two primary objectives are addressed. The first is to convert the watermark format for more effective feature extraction. The second is to directly encode the temporal-domain waveform and watermark, thereby generating the watermarked signal. To achieve the first objective, the dense block $\mathbf{DB}$ is designed, whereas the integrated content-driven encoder (ICDE) $\mathbf{E}$ is constructed to fulfill the second. The detailed architectures and watermarking process are presented as follows. 

\textit{Architectures}: The dense block consists of two fully connected (FC) layers interleaved with a ReLU activation function, enabling it to process the watermark of varying lengths effectively.
MobileNetV2~\cite{sandler2018mobilenetv2} indicates that ReLU-like activation functions can result in substantial information loss, particularly for low-dimensional features. Given the critical role of low-dimensional features in waveform reconstruction, the proposed ICDE architecture eschews normalization layers and activation functions. Instead, it employs a downsampling block consisting of four one-dimensional convolutional (Conv1d) layers, followed by an upsampling block with four transpose Conv1d layers and four additional Conv1d layers. This fully convolutional design not only preserves low-dimensional features but also ensures the lightweight nature of the ICDE.

%The ICDE includes a downsampling block composed of four convolutional layers. Additionally, it features an upsampling block consisting of four transpose convolutional layers and is complemented by four convolutional layers. All convolutional layers are one-dimensional. The lightweight structure is attained due to the ICDE comprising solely convolutional layers. 

%The exclusion of normalization layers and activation functions stems from the observation in MobileNetV2~\cite{sandler2018mobilenetv2} that activation functions akin to ReLU can result in substantial information loss for low-dimensional features. 

%As the encoder operates exclusively on low-dimensional features during waveform reconstruction, we have eschewed complex network architectures to maximize the preservation of original waveform information.

\textit{Watermarking Process}: 
Given a watermark $\mathbf{w} \in \{0, 1\}^l$, where $l$ is the length of the watermark.  $\mathbf{DB}$ is utilized to transform the watermark into the latent variable $\mathbf{\sigma_w}$:
\begin{equation}
\setlength{\abovedisplayskip}{0.1cm}
\setlength{\belowdisplayskip}{0.1cm}
    \mathbf{\sigma_w} = \mathbf{DB}(\mathbf{w}) \in \mathbb{R}^{B \times C \times L},
\end{equation}
where $B$ represents the batch size, $C$ denotes the channel of the waveform, and $L$ is the length of the waveform.
Then, the latent variable $\mathbf{\sigma_w}$ is concatenated with the waveform $\mathbf{s}$ to acquire the final input $\sigma$ of the encoder:
\begin{equation}
\setlength{\abovedisplayskip}{0.1cm}
\setlength{\belowdisplayskip}{0.1cm}
    \mathbf{\sigma} = \mathbf{s} \oplus \mathbf{\sigma_w} \in \mathbb{R}^{B \times C \times L},
\end{equation}
where $\oplus$ represents channel concatenation. The encoder $\mathbf{E}(\cdot)$ takes $\sigma$ as input to reconstruct the watermarked waveform $\hat{\mathbf{s}}$.
The complete watermarking process can be formalized as:
\begin{equation}
\setlength{\abovedisplayskip}{0.1cm}
\setlength{\belowdisplayskip}{0.1cm}
    \hat{\mathbf{s}} = \mathbf{E}(\mathbf{s} \oplus \mathbf{DB}(\mathbf{w})).
\end{equation}

% \textcolor{red}{Have revised.}

% -----------------------------------------------------------
\vspace{-0.45cm}
\subsection{Extracting Module}
\label{sec_em}
The extracting module consists of a convolutional block (ConvBlock) $\mathbf{D}$ and an extraction block $\mathbf{EB}$. The $\mathbf{D}$ aims to isolate the watermark features from the watermarked waveform, while the $\mathbf{EB}$ reconstructs the extracted watermark. Specific architectures of the module are available here.

\textit{Architectures}: Gating mechanism\cite{jozefowicz2016exploring} has demonstrated its effectiveness in feature extraction for language processing tasks\cite{dauphin2017language} and classification tasks\cite{yu2024mambaout}. To fully exploit this advantage, we have designed the gating mechanism to capture temporal domain features, resulting in a Temporal-aware Gated Convolutional Network (TGC), which incorporates the ConvBlock in the extracting module. The TGC follows a dual-branch structure, where the main branch consists of a sequential arrangement of a Conv1d layer, batch normalization, and a sigmoid activation to implement the gating mechanism. In contrast, the shortcut branch contains only a Conv1d layer for acquiring fine-grained features from the input. The outputs of these two branches are merged using the Hadamard Product to produce the final output. This efficient design contributes to the lightweight nature of the ConvBlock. 

\textit{Attack simulator}: As outlined in the Introduction, ensuring strong robustness relies heavily on the attack simulator (AS), which serves as a crucial component of the overall architecture (illustrated in Fig.~\ref{fig_archi}). Hence, the AS also plays an integral role in the effectiveness of our proposed method by incorporating nine voice post-processing operations: Gaussian noise (GN), low-pass filtering (LP), band-pass filtering (BP), high-pass filtering (HP), time stretching and interpolation (TSI), suppression (SPS), resampling (ReS), echo, and dither. In practice, the AS is utilized exclusively during the training phase, with a single attack applied to each waveform. 

%The ConvBlock is composed of seven gated convolutional networks~\cite{dauphin2017gcn}, and the extraction block consists of two FC layers and a ReLU activation. Owing to the superior feature extraction capabilities of gated convolutions, they assume a pivotal role in classification tasks, as exemplified by \textit{MambaOut}~\cite{yu2024mambaout}. Consequently, we redesigned a Temporal-aware Gated Convolutional Network (TGC), which retains a dual-branch structure. One branch comprises a single one-dimensional convolutional layer for extracting the original features of input. In the alternate branch, features initially undergo batch normalization following an identical convolutional operation, subsequently achieving a gating mechanism via a Sigmoid activation. The outputs of both branches are then merged by Hadamard Product to generate the final output. Moreover, the extraction block consists of a FC layer, a ReLU activation, and a FC layer.

\textit{Extracting \& Verification}: 
The procedure for extracting the watermark $\hat{\mathbf{w}}$ can be formalized as follows:
\begin{equation}
\setlength{\abovedisplayskip}{0.1cm}
\setlength{\belowdisplayskip}{0.1cm}
    \hat{\mathbf{w}} = \mathbf{EB}(\mathbf{D}(\hat{\mathbf{s}})).
\end{equation}

\noindent Inspired by \cite{lin2022cycleganwm},  test hypothesis is employed for watermark verification. Assuming that the errors in the watermark bits are independent and taking into account the previously defined watermark bit length $l$, the number of matching watermark bits $\kappa$ is calculated using the binomial distribution 
$Pr(X=\kappa) = \sum_{i=\kappa}^l \binom{l}{i} \xi^i(1-\xi)^{l-i}$, 
where $\xi=0.5$ is the probability that needs to be tested under hypotheses.

\begin{table}[]
\centering
\renewcommand\arraystretch{0.9}
\setlength{\abovecaptionskip}{-0.01cm}
\caption{Robustness results with/without attack simulator.}
\resizebox{0.95\linewidth}{!}{
\begin{tabular}{ccccccc}
\toprule
 & Method & GN 10dB & GN 20dB & LP 3k  & BP 0.8-5k & TSI 2$\times$     \\ 
 \midrule
\multicolumn{7}{l}{TAWM~\cite{liu2023timebre}} \\
 & \multicolumn{1}{r}{w/o. AS} & 0.6012     & 0.7849    & 0.9176 & 0.9129 & 0.8348 \\
 & \multicolumn{1}{r}{w. AS} & 0.6335     & 0.8154    & 0.9934 & 0.9983 & 0.9448 \\ 
 \midrule
\multicolumn{7}{l}{True (Ours)} \\
 & \multicolumn{1}{r}{w/o. AS} & 0.4994     & 0.5314    & 0.4960 & 0.6087 & 0.5006 \\
 & \multicolumn{1}{r}{w. AS} & 0.7922     & 0.9358    & 0.9096 & 0.9687 & 0.9678 \\ 
 \bottomrule
\end{tabular}
}
\label{tab_woas}
\vspace{-0.2cm}
\end{table}

\begin{table}[t]
\vspace{-0.2cm}
    \centering
    \setlength{\abovecaptionskip}{-0.01cm}
    \setlength{\belowcaptionskip}{-0.01cm}
    \caption{Comparison of fidelity with handcraft (HC) watermarking methods and deep-learning-based (DL) methods.}
    \resizebox{0.95\linewidth}{!}{
    \begin{tabular}{ccccccc}
    \toprule
           & Method (bps) & Dataset     & STOI↑           & PESQ↑           & SSIM↑           & ACC↑            \\
           \midrule
           \multirow{3}{*}{HC}& FSVC (32)\cite{zhao2021desyn-fsvcm}  & \multirow{3}{*}{LJSpeech}    & 0.9984          & 3.9977          & 0.9803          & \textbf{1.0000}          \\
           & Normspace (32)\cite{saadi2019normspace}  &     & 0.9646          & 2.5506          & 0.8868          & \textbf{1.0000}          \\
           & PBML (100)\cite{natgunanathan2017patchwork}  &     & 0.9861          & 3.7866          & 0.9560          & \textbf{1.0000}          \\
           \cdashline{1-7}[2pt/2pt]
             \multirow{4}{*}{DL}& AudioSeal (16)\cite{roman2024audioseal}  & \multirow{4}{*}{LJSpeech}    & 0.9985          & 4.5893          & 0.9811          & 0.9214          \\
            & WavMark (32)\cite{chen2023wavmark}    &     & \textbf{0.9997} & 4.4628          & 0.9690          & \textbf{1.0000} \\
            & TAWM (100)\cite{liu2023timebre}       &     & 0.9853          & 4.0353          & 0.9388          & 0.9998          \\
            & GROOT (100)\cite{liu2024groot}       &     & 0.9605        & 
            3.3871          & 0.9088       &0.9969\\
           \cdashline{2-7}[2pt/2pt]
            \rowcolor[HTML]{DDDDDD}
            & True (32) (Ours) & LJSpeech    & 0.9986          & 4.5748 & \textbf{0.9833} & 0.9986          \\
            \rowcolor[HTML]{DDDDDD}
            & True (100) (Ours) & LJSpeech    & 0.9987          & \textbf{4.6380} & 0.9819 & 0.9973          \\
            \midrule
            \rowcolor[HTML]{DDDDDD}
            & True (100) (Ours) & LibriTTS    & 0.9967          & 4.6290          & 0.9889          & 1.0000          \\
            \rowcolor[HTML]{DDDDDD}
            & True (100) (Ours) & LibriSpeech & 0.9985          & 4.6218          & 0.9753          & 0.9992          \\
            \rowcolor[HTML]{DDDDDD}
            & True (100) (Ours) & M4Singer    & 0.9799          & 4.6193          & 0.9939          & 0.9992          \\
            \rowcolor[HTML]{DDDDDD}
            & True (100) (Ours) & Opencpop    & 1.0000          & 4.6400          & 0.9971          & 0.9978          \\
            \bottomrule
    \end{tabular}
    }
    \label{tab_fidelity}
    \vspace{-0.65cm}
\end{table}

% -----------------------------------------------------------
\vspace{-0.4cm}
\subsection{Optimizing Strategy}
\vspace{-0.1cm}
\label{sec_train}
% During training, we endeavor to enhance watermark extraction accuracy while preserving the watermarked waveform fidelity by concurrently optimizing the HM, AS, and EM.
The ultimate objective of jointly optimizing the HM, AS, and EM is to balance the trade-off between watermark extraction accuracy and the fidelity of the watermarked voice.
To ensure high fidelity, the mel-spectrogram loss is initially employed to minimize the distance between the natural waveform $\mathbf{s}$ and the watermarked waveform $\hat{\mathbf{s}}$, which is formulated as:
\begin{equation}
\setlength{\abovedisplayskip}{0.1cm}
\setlength{\belowdisplayskip}{0.1cm}
    \mathcal{L}_{MEL} = || \psi(\mathbf{s}) - \psi(\hat{\mathbf{s}}) ||_1,
\end{equation}
where $||\cdot||_1$ is $L_1$ norm and $\psi$ represents the function of mel transformation. Then, the logarithmic STFT magnitude loss is utilized as an additional measure to further enhance fidelity:
\begin{equation}
\setlength{\abovedisplayskip}{0.1cm}
\setlength{\belowdisplayskip}{0.1cm}
    \mathcal{L}_{MAG} = || \log(\mathbf{STFT}(\mathbf{s})) - \log(\mathbf{STFT}(\hat{\mathbf{s}})) ||_1,
\end{equation}
where $\mathbf{STFT}(\cdot)$ denotes the transformation of STFT magnitude. The total loss for preserving the fidelity is defined as:
\begin{equation}
\setlength{\abovedisplayskip}{0.1cm}
\setlength{\belowdisplayskip}{0.1cm}
    \mathcal{L}_{WAV} = \lambda_{1} \mathcal{L}_{MEL} + \lambda_{2} \mathcal{L}_{MAG},
\end{equation}
where $\lambda_{1}$ and $\lambda_{2}$ are hyper-parameters of the mel-spectrogram loss and the logarithmic STFT magnitude loss.

Binary cross-entropy is leveraged to ensure the accurate extraction of the watermark:
\begin{equation}
\setlength{\abovedisplayskip}{0.1cm}
\setlength{\belowdisplayskip}{0.1cm}
    \mathcal{L}_{WM} = - \sum_{i=1}^{k} w_i \log {\hat{w}_i} + (1-w_i) \log(1-\hat{w}_i).
\end{equation}

The overall training loss is formulated as follows:
\begin{equation}
\setlength{\abovedisplayskip}{0.1cm}
\setlength{\belowdisplayskip}{0.1cm}
    \mathcal{L} = \mathcal{L}_{WAV} + \alpha \mathcal{L}_{WM},
\end{equation}
where $\alpha$ serves a hyper-parameter to balance auditory quality and watermark recovery accuracy.

%\begin{figure}[t]
%    \centering
%    \setlength{\abovecaptionskip}{-0.01cm}
%    \setlength{\belowcaptionskip}{-0.01mm}
%   \resizebox{\linewidth}{!}{\includegraphics{fig/comparison.pdf}}
%    \caption{Comparison of Fidelity With Other SOTA Methods.}
%    \label{fig_comparison_fidelity}
%\vspace{-0.4cm}
%\end{figure}

\begin{table}[t]
        \setlength{\abovecaptionskip}{-0.01cm}
        \setlength{\belowcaptionskip}{-0.01cm}
        \centering
        \renewcommand\arraystretch{0.9}
        \caption{Capacity of the proposed method under various datasets.}
        \resizebox{0.95\linewidth}{!}{
        \begin{tabular}{ccccccc}
            \toprule
            \multirow{2}{*}{Dataset}&  & \multicolumn{5}{c}{Capacity (\textit{bps})} \\[-0.5 ex]
            \cmidrule{3-7}
            & & 32     & 100    & 300    & 500    & 600    \\[-0.5 ex]
            \midrule
            \multirow{2}{*}{LJSpeech} & PESQ↑ & 4.5748 & 4.6380 & 4.6429 & 4.6000 & 4.6200 \\
             & ACC↑  & 0.9997 & 0.9973 & 0.9956 & 0.9210 & 0.8985 \\
            \cmidrule{1-7}
            \multirow{2}{*}{Opencpop} & PESQ↑ & 4.5598 & 4.6400 & 4.6137 & 4.2753 & 4.6227 \\
             & ACC↑  & 0.9873 & 0.9978 & 0.9735 & 0.9716 & 0.8956 \\
            \bottomrule
        \end{tabular}
        }
        \label{tab_capacity}
\vspace{-0.5cm}
\end{table}

\vspace{-0.2cm}
\section{Experimental Results and Analysis}
\vspace{-0.1cm}
\label{sec_method}
%In this section, we conduct comprehensive experiments to rigorously evaluate our proposed method across several dimensions: fidelity, capacity, and robustness. In addition, we undertake a comparative analysis of our approach against the current state-of-the-art (SOTA) methods. The details of these experiments and their respective analyses are elaborated below.

This section presents experiments conducted to evaluate the proposed True method based on three primary assessment criteria: fidelity, capacity, and robustness. Additionally, comprehensive experiments and analyses are provided to compare its performance against SOTA methods, further demonstrating its effectiveness. 

% -----------------------------------------------------------
\vspace{-0.3cm}
\subsection{Experimental Setup}
\vspace{-0.1cm}
\label{sec_es}
\textit{Datasets and Baseline}: In the context of voice datasets, other than conventional speech voice datasets like LJSpeech~\cite{Ito2017ljspeech}, LibriTTS~\cite{zen2019libritts}, and Aishell3~\cite{shi2020aishell3}, singing voice datasets like M4Singer~\cite{zhang2022m4singer} and Opencpop~\cite{wang2022opencpop} are also considered. 
%For our experiments, we utilized three speech voice datasets: LJSpeech~\cite{Ito2017ljspeech}, LibriTTS~\cite{zen2019libritts}, and Aishell3~\cite{shi2020aishell3}, in addition to two singing voice datasets: M4Singer~\cite{zhang2022m4singer} and Opencpop~\cite{wang2022opencpop}. 
% LJSpeech and LibriTTS are single-speaker and multi-speaker English datasets. Aishell3 is a multi-speaker Chinese dataset. M4Singer and Opencpop are single-speaker and multi-speaker Chinese singing voice datasets.
Moreover, comprehensive comparisons against SOTA watermarking methods are conducted, including handcrafted (HC) methods such as FSVC~\cite{zhao2021desyn-fsvcm}, Normspace~\cite{saadi2019normspace}, and PBML~\cite{natgunanathan2017patchwork}, as well as deep learning-based (DL) methods like AudioSeal~\cite{roman2024audioseal}, WavMark~\cite{chen2023wavmark}, TAWM~\cite{liu2023timebre}, and Groot~\cite{liu2024groot}.

\textit{Evaluation Metrics}: Two auditory objective metrics, Short-Time Objective Intelligibility (STOI)~\cite{taal2010short} and Perceptual Evaluation of Speech Quality (PESQ)~\cite{recommendation2001perceptual}, are used to evaluate the auditory performance of watermarked voice. Besides, the Structural Similarity Index Measure (SSIM)~\cite{wang2004image} is employed to assess the watermarked voice from the perspective of the visualized spectrogram. 

\textit{Models \& Training Settings}: In the downsampling block of ICDE, Conv1d layers use a kernel size of 3, a stride of 2, and padding of 2, except for the last layer, which has padding of 1. In the upsampling block, Conv1d layers maintain a kernel size of 3, a stride of 1, and padding of 1, while transposed Conv1d layers have a kernel size of 3, a stride of 2, padding of 2, and output padding of 1, except for the first layer, which uses padding of 1. For the TGCs, both Conv1d layers in the two branches have a kernel size of 3, a stride of 2, and padding of 1. 
During training, AdamW optimizer~\cite{loshchilov2018adamw} is employed with a learning rate of 2e-4. The training is conducted for 40 epochs with a batch size of 16. The hyper-parameters $\lambda_1$, $\lambda_2$, and $\alpha$ are set to 0.8, 0.1, and 0.3, respectively.

%\textit{AS Settings}: 
% ----table--------------------------------------------------
\begin{table*}[t]
\centering
\renewcommand\arraystretch{0.9}
\setlength{\abovecaptionskip}{-0.01cm}
\caption{Robustness of the proposed method in terms of accuracy under various datasets.}
\resizebox{0.95\textwidth}{!}{
\begin{tabular}{cccccccccccccc}
    \toprule
     & \multirow{2}{*}{Dataset} & \multicolumn{3}{c}{GN} & PN & LP & BP & HP & SPS & ReS & Echo & TSI & Dither   \\
    \cmidrule{3-14}
     & & 10 dB & 15 dB & 20 dB & 0.5 & 3k & 0.5-8k & 1k & behind & 44.1k & default & 0.5$\times$ & default \\
     \midrule
    \multirow{2}{*}{\makecell{Singing \\ Voice}}& M4Singer & 0.9254 & 0.9783 & 0.9961 & 0.9645 & 0.9903 & 0.9992 & 0.9841 & 0.9992 & 0.9992 & 0.9896 & 0.9992 & 0.9992  \\
    & Opencpop & 0.8260 & 0.9298 & 0.9782 & 0.9625 & 0.8305 & 0.9954 & 0.9872 & 0.9960 & 0.9961 & 0.9802 & 0.9940 & 0.9959  \\
    \midrule
    \multirow{3}{*}{\makecell{Speech \\ Voice}}& LJSpeech & 0.7922 & 0.8844 & 0.9358 & 0.9964 & 0.9096 & 0.9687 & 0.8417 & 0.9679 & 0.9699 & 0.9216 & 0.9678 & 0.9970  \\
    & LibriTTS & 0.8285 & 0.9117 & 0.9557 & 1.0000 & 0.9471 & 0.9857 & 0.9084 & 0.9856 & 0.9857 & 0.9396 & 0.9852 & 1.0000  \\
    & Aishell3 & 0.9257 & 0.9763 & 0.9935 & 0.9992 & 0.8919 & 0.9991 & 0.9978 & 0.9991 & 0.9992 & 0.9876 & 0.9990 & 0.9992  \\
    \bottomrule
\end{tabular}
}
\label{tab_robust}
\vspace{-0.6cm}
\end{table*}

\begin{table}[]
\centering
\setlength{\abovecaptionskip}{-0.01cm}
\caption{Comparison of robustness with SOTA methods in accuracy.}
\resizebox{0.98\linewidth}{!}{ 
\begin{tabular}{ccccccc}
    \toprule
    \multirow{2}{*}{Method (bps)} & \multicolumn{2}{c}{GN} & PN & BP & SPS & Echo \\[-0.5 ex]
    \cmidrule{2-7}
     & 10 dB & 20 dB & 0.5 & 0.5-8k & behind & default \\[-0.5 ex]
     \midrule
    AudioSeal (16)~\cite{roman2024audioseal} & 0.6086 & 0.6600 & 0.6571 & 0.9764 & 0.8925 & 0.7277  \\
    Normspace (32)~\cite{saadi2019normspace} & 0.5856 & 0.6208 & 0.4733  & 0.4796 & 0.6596 & 0.5630  \\
    FSVC (32)~\cite{zhao2021desyn-fsvcm} & 0.7312 & 0.8835 & 0.8164 & 0.8263 & 0.7188 & 0.7976  \\
    WavMark (32)~\cite{chen2023wavmark} & 0.5295 & 0.6523 & 0.6924 & 0.9995 & 0.9713 & 0.8668  \\
    \rowcolor[HTML]{DDDDDD}
    \textbf{True (32) (Ours)} & \textbf{0.9673} & \textbf{0.9986} & \textbf{0.9868} & \textbf{0.9998} & \textbf{0.9996} & \textbf{0.9963} \\
    \midrule
    PBML (100)~\cite{natgunanathan2017patchwork} & 0.6060 & 0.7176 & 0.7060 & 0.7504 & 0.6365 & 0.6995  \\
    TAWM (100)~\cite{liu2023timebre} & 0.6335 & 0.8154 & 0.7282 & \textbf{0.9983} & 0.9814 & 0.9471 \\
    Groot (100)~\cite{liu2024groot}  &\textbf{0.9929} &  \textbf{0.9953} & 0.9861 & 0.9947 & 0.9718 & \textbf{0.9833} \\
    \rowcolor[HTML]{DDDDDD}
    \textbf{True (100) (Ours)} & 0.8285& 0.9557 & \textbf{0.9964} & 0.9857 & \textbf{0.9856} & 0.9396  \\
    \bottomrule
\end{tabular}
}
\label{tab_robust_comparison}
\vspace{-0.6cm}
\end{table}

%All experiments are carried out on a single NVIDIA GeForce RTX 3090 GPU. 
% During the training process, the hyper-parameters $\lambda_{mel}$ and $\lambda_{mag}$ are 0.9 and 0.1 respectively. And we initially set the hyper-parameters $\alpha$ to 1. When $\mathcal{L}_{WM}$ reaches the predetermined threshold, $\alpha$ is adjusted to 0.1.

\vspace{-0.3cm}
\subsection{Indispensable AS for Frequency-domain Watermarking }
\vspace{-0.1cm}
The observation presented in the Introduction section highlights the significance of AS in frequency-domain watermarking techniques. To validate this, we select the TAWM~\cite{liu2023timebre}, which serves as a representative example of frequency-domain watermarking methods for our analysis.
%To validate our observation that the robustness depends on the attack simulator for frequency-domain watermarking, we examined the TAWM~\cite{liu2023timebre} as a case study.
Table~\ref{tab_woas} clearly demonstrates the difference in robustness of TAWM with and without the AS when subjected to various attacks.
%the ensuing robustness is lower than that attained with AS. 
%Particularly in scenarios with a high watermark extraction accuracy, the absence of AS participation results in a performance reduction of nearly 8\%.
In the absence of AS, extraction accuracy can decrease by nearly 8\% under certain attacks.
%These findings substantiate the critical insight that, while frequency-domain features offer a preliminary level of robustness, incorporating an attack simulator is essential for achieving robust watermarking performance.
These findings reaffirm that, while frequency-domain watermarking offers robust protection, the inclusion of AS is crucial for achieving a higher level of robustness.

% -----------------------------------------------------------
\vspace{-0.4cm}
\subsection{Fidelity and Capacity}
\textit{Fidelity} refers to the imperceptibility of embedded watermarks, measured by the extent to which audio quality is preserved after watermarking. Table~\ref{tab_fidelity} reports both audio quality (for speech and singing voice) and the corresponding watermark extraction accuracy achieved by the proposed True. Additionally, it presents a fidelity comparison against SOTA methods on the LJSpeech dataset. All metrics are computed by comparing the watermarked waveform to its natural counterpart. The results indicate that True achieves excellent speech quality while maintaining reasonable watermark extraction accuracy on speech datasets. Similarly, it demonstrates strong fidelity and extraction performance on singing voice datasets. In comparative experiments with SOTA methods, True ranks first in both PESQ and SSIM metrics, highlighting its superior speech quality. 
% Although its extraction accuracy did not reach a perfect 100\%, our method still achieved an impressive 99.73\% accuracy, which falls within the reasonable range validated in Section~\ref{sec_em}.

\textit{Capacity} which reflects the length of watermarks that can be embedded, is as critical as fidelity in evaluating watermarking performance. Table~\ref{tab_capacity} illustrates the results of the proposed method on the LJSpeech and Opencpop datasets under varying watermark capacities. 
Experimental results demonstrate that True is well-suited for high-capacity scenarios, supporting up to 500 bps. On the LJSpeech dataset, the method achieves an average watermark extraction accuracy of 97.84\% while maintaining high fidelity, with an average PESQ score of 4.6140. Similarly, for the Opencpop dataset, it achieves an average PESQ of 4.5222 and a recovery accuracy of 98.26\%. However, when the capacity exceeds 600 bps, both fidelity and extraction accuracy degrade significantly.

% -----------------------------------------------------------
\vspace{-0.3cm}
\subsection{Robustness}
We evaluated the robustness of the proposed method across various datasets at a watermark capacity of 100 bps. GN, Pink noise (PN), LP, BP, HP, SPS, ReS, Echo, TS, and Dither were employed for validation. Table~\ref{tab_robust} illustrates the watermark extraction accuracy under these attacks for each dataset.
The results show that the proposed method exhibits strong robustness, particularly on singing voice datasets. Specifically, it achieves extraction accuracies of 99.61\% and 97.82\% under a noise level of 20 dB. Under remanent attacks, True further attains average recovery accuracies of 99.45\% and 96.85\%, respectively.
Although performance on speech datasets is slightly lower than that on singing voice, True still maintains desirable robustness, with average extraction accuracies of 92.33\%, 94.85\%, and 91.91\% across the evaluated conditions.

The proposed True was further compared with SOTA methods on the LJSpeech dataset to evaluate its robustness against various voice post-processing operations, %including GN, PN, BP, SPS, and Echo.
%The extraction accuracies following post-processing are
as illustrated in Table~\ref{tab_robust_comparison}.
To account for the capacity limitation of the compared methods, two sets of experiments were conducted at capacities of 32 bps and 100 bps, respectively.
At 32 bps, True achieved higher watermark extraction accuracy than all SOTA methods. Notably, it exhibited strong resilience, maintaining an average extraction accuracy of 99.14\% even after undergoing six different types of signal attacks.
%Under this capacity, the robustness surpassing that of SOTA methods can be attributed to ICDE and TGC in True.
%While conventional attacks target only surface-level speech signals, ICDE embeds watermarks into high-level temporal features that are harder to manipulate.
%Furthermore, TGC leverages its gating mechanism to filter out shallow features, enabling the disentangling of watermark features from high-level features for reliable watermark extraction.
The observed robustness gains can be attributed to ICDE's deep feature embedding and TGC's gated decoding, which jointly preserve and extract watermark signals from high-level temporal features. This design makes the method more resilient to typical signal-level distortions.
Under the 100 bps configuration, True continued to demonstrate robust performance, particularly in resisting PN and SPS attacks. Although its accuracy under 10 dB GN was slightly lower than that of Groot, True still significantly outperformed the remaining two methods.

% -----------------------------------------------------------
\vspace{-0.17cm}
\section{Conclusion}
\vspace{-0.13cm}
In this study, we propose \textit{True}, a temporal-aware robust watermarking method designed to proactively protect the copyrights of diverse waveform types, including both speech and singing voice.
To enable seamless watermark embedding, we introduce a content-driven encoder that directly integrates the watermark into the temporal representation of the waveform and reconstructs the watermarked signal end-to-end.
For extraction, we develop a temporal-aware gated convolutional network that effectively captures fine-grained features from attacked waveforms, thereby enhancing watermark recovery accuracy. The proposed True surpasses baseline methods in fidelity, supports high-capacity embedding of up to 500 bps, and exhibits strong robustness against a wide range of common waveform distortions.

% In this paper, we propose True, a universal framework for proactively protecting the copyright of speech and singing voices via the temporal-aware robust watermarking method. 
% To this end, we meticulously design a hiding module, comprising a structure-lightweight intergraded content-driven encoder, for watermarking in time-domain features and reconstruction of the watermarked waveform. 
% In the extracting module, we further devise a temporal-aware gated convolution network that effectively captures high-dimensional temporal characteristics of the waveform, thereby enabling more precise watermark recovery. 
% Comprehensive validation experiments and comparisons with state-of-the-art methods further corroborate the superior fidelity and powerful robustness against common voice post-processing operations of True.

%----------------------------------------------------------------------
% \newpage
% \clearpage
\bibliographystyle{IEEEtran}
\bibliography{ref.bib}

\end{document}